\documentclass{article}
\usepackage{spconf,amsmath,graphicx,hyperref}
\usepackage{cite}
\usepackage{bm}
\usepackage{multicol}
\usepackage{pgfplots}
\usetikzlibrary{shadows, decorations.pathmorphing, shapes.geometric, arrows.meta}
\usepackage{subcaption}
\pgfplotsset{compat=newest}
\pgfplotsset{plot coordinates/math parser=false}
\usepackage{amsmath,amssymb,amsfonts}
\usepackage{algpseudocode}
\usepackage{textcomp}
\usepackage{xcolor}
\usepackage{upgreek}
\usepackage{algorithm}
\usepackage{booktabs}
\begin{document}

\title{Hardware-Efficient Cognitive Radar: Multi-Target Detection with RL-Driven Transmissive RIS}
\name{Adam Umra$^{\star}$, Aya Mostafa Ahmed$^{\dagger}$, Stefan Roth$^{\star}$, Aydin Sezgin$^{\star}$%
\thanks{This work was supported in part by the German Research Foundation (“Deutsche Forschungsgemeinschaft”) (DFG) under Project–ID 287022738 TRR 196 for Project S03 and in part by the German Federal Ministry of Research, Technology and Space (BMFTR) in the course of IGEL AI under grant 16KIS2342.}}

\address{$^{\star}$ Ruhr University Bochum, Germany \\
         $^{\dagger}$ Bosch GmbH, Germany}

\maketitle
\begin{abstract}
Cognitive radar has emerged as a key paradigm for next-generation sensing, enabling adaptive, intelligent operation in dynamic and complex environments. Yet, conventional cognitive multiple-input multiple-output (MIMO) radars offer strong detection performance but suffer from high hardware complexity and power demands. To overcome these limitations, we develop a reinforcement learning (RL)-based framework that leverages a transmissive reconfigurable intelligent surface (TRIS) for adaptive beamforming. A state–action–reward–state–action (SARSA) agent tunes TRIS phase shifts to improve multi-target detection in low signal-to-noise ratio (SNR) conditions while operating with far fewer radio frequency (RF) chains. Simulations confirm that the proposed TRIS–RL radar matches or, for large number of elements, even surpasses MIMO performance with reduced cost and energy requirements.
\end{abstract}


\section{Introduction}
\label{sec:intro}
Cognitive radar (CR) is \, an emerging radar paradigm that adapts its operation through a perception–action cycle. Unlike fixed sensing strategies, CR observes the environment, learns from it, and adjusts decisions in real time. This closed-loop adaptability is crucial in nonstationary settings—such as noise, clutter, or unknown targets—where traditional radars struggle. By embedding learning and feedback into sensing, CR achieves significant performance gains over conventional, non-adaptive systems~\cite{gurbuz2019CR,haykin2012CR}.
While CR provides the structural foundation for adaptive sensing, its full potential is realized when paired with advanced decision-making frameworks. In particular, reinforcement learning (RL) offers the means to systematically learn and refine sensing strategies through direct interaction with the environment~\cite{sutton2017RLIntro}. By continuously updating its actions based on feedback, in the RL terminology referred to as rewards, RL optimizes how targets are detected and tracked—without needing detailed statistical models of the environment. This makes RL-driven CR systems more flexible and resilient~\cite{ahmed2021RLradar}.
Multiple-input multiple-output (MIMO) radar systems have been extensively studied in CR applications due to their inherent ability to enhance performance through waveform diversity and spatial multiplexing~\cite{ahmed2021RobustRL,umra2025smartersensing2dclutter,fortunati2023ISAC}. However, the associated hardware complexity, increased power consumption, and physical size of multiple active radio frequency (RF) chains present substantial challenges, particularly in scenarios demanding compact size or stringent energy constraints.

To solve these challenges, this paper introduces a architecture leveraging a passive transmissive reconfigurable intelligent surface (TRIS) coupled to a single-antenna transmitter~\cite{zhu2025TRIS}. The TRIS functions as a spatial waveform modulator, dynamically reshaping radiation patterns without numerous active RF components, thereby reducing hardware complexity and power requirements. A uniform planar array (UPA) receiver complements the TRIS-based transmitter, enabling precise beamforming and high-resolution target localization, ensuring a balanced trade-off between transmitter simplicity and receiver performance. We develop a mathematical framework by modeling the TRIS in the nearfield of the transmit antenna, deriving the signal and detection models, and formulating target detection as a binary hypothesis test with a constant false alarm rate (CFAR) adaptive matched filter. Adaptivity is achieved via a SARSA-based reinforcement learning agent that optimizes TRIS phase configurations in a state--action--reward cycle. Monte Carlo simulations benchmark the proposed radar against conventional MIMO systems, showing superior low-SNR multi-target detection with reduced hardware complexity.
\section{SYSTEM MODEL}
\begin{figure}[t]
  \centering
  \centerline{\includegraphics[width=\linewidth]{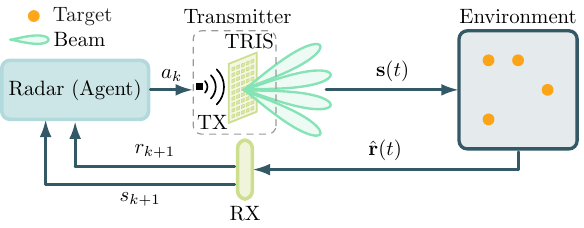}}
  \caption{RL-Based TRIS-Enabled CR for Enhanced Multitarget Detection}\label{fig:system}
\end{figure}
We consider a colocated radar system comprising a single transmit antenna and $N_\mathrm{R}$ receive antennas arranged as a uniform planar array (UPA). The receive array has an inter-element spacing of $\lambda/2$, where $\lambda$ denotes the wavelength corresponding to the operating frequency. The transmitter is augmented with a TRIS, which consists of $N$ radiating elements organized in a UPA configuration. An illustration of the system is provided in Figure~\ref{fig:system}.
\subsection{TRIS Model}
The TRIS imparts a controllable phase shift onto the transmitted signals. This effect is captured by a diagonal phase shift matrix $\mathbf{\Phi} \in \mathbb{C}^{N \times N}$ given by $\mathbf{\Phi} = \text{diag}(e^{j\varphi_1}, e^{j\varphi_2}, \dots, e^{j\varphi_N})$, where $\varphi_n \in [0, 2\pi)$ denotes the phase shift applied by the $n$-th TRIS element. Signal propagation between the transmit antenna and the $N$ TRIS elements is described by a transmission vector $\mathbf{w} = [w_1,\dots,w_N]^T \in \mathbb{C}^{N\times 1}$. Since the TRIS is in the near-field of the antenna, each element $w_{n}$ is characterised based on the Rayleigh–Sommerfeld diffraction theory~\cite{an2024SIM}
\begin{equation}
w_{n} = \frac{d_x d_y \cos(\alpha_{n})}{d_{n}^2} \left(\frac{1}{2\pi d_{n}} - j\frac{1}{\lambda}\right) e^{j 2\pi d_{n} / \lambda},
\end{equation}
where $d_{n}$ is the distance from the transmit antenna to the $n$-th TRIS element, $\alpha_{n}$ is the angle between the propagation path and the surface, and $d_x \times d_y$ represents the physical size of each TRIS element. $d_n$ and $\alpha_{n}$ can be computed from geometry, as described in~\cite{an2025SIM}.
\subsection{Signal Model}
The transmit signal vector is
\begin{equation}
s(t) = \sqrt{P_\mathrm{T}}\psi(t) \in \mathbb{C},
\end{equation}
where $\psi(t)$ comprises a orthonormal waveform, and $\sqrt{P_\mathrm{T}}$ is the transmit power. Let $\mathbf{g} = \mathbf{\Phi} \mathbf{w} \in \mathbb{C}^{N\times 1}$ denote the effective transmission vector incorporating TRIS effects. The received baseband signal, reflected from a single point-like target located at azimuth $\theta$ and elevation $\phi$, is modeled as
\begin{equation}
\hat{\mathbf{r}}(t) = \alpha \mathbf{a}_\mathrm{R}(\theta, \phi) \mathbf{a}_\mathrm{T}^T(\theta, \phi) \mathbf{g} s(t - \tau) + \hat{\mathbf{n}}(t),
\end{equation}
where $\alpha \in \mathbb{C}$ captures the radar cross-section and two-way path loss (modeled by the Swerling 0 model~\cite{swerling1960Prob}), and $\tau$ is the propagation delay. The vectors $\mathbf{a}_\mathrm{T}(\theta, \phi) \in \mathbb{C}^{N\times 1}$ and $\mathbf{a}_\mathrm{R}(\theta, \phi) \in \mathbb{C}^{N_\mathrm{R}\times 1}$ are the transmit and receive steering vectors. The disturbance term $\hat{\mathbf{n}}(t)$ accounts for clutter and additive white Gaussian noise. The beamforming pattern at the TRIS can be expressed as $B(\theta, \phi) = \mathbf{a}_\mathrm{T}^\text{T}(\theta, \phi) \mathbf{G} \mathbf{G}^\text{H} \mathbf{a}_\mathrm{T}^*(\theta, \phi)$.
\subsection{Matched Filtering}
At the receiver, the signal is processed through a matched filter aligned to a delay estimate $\hat{\tau}$. The output of the matched filter is given by
\begin{equation}
\mathbf{r}(\hat{\tau}) = \int_0^T \hat{\mathbf{r}}(t) \psi_{\text{MF}}^\text{*}(t - \hat{\tau}) \,\mathrm{d}t.
\end{equation}
Substituting the received signal expression, we obtain
\begin{align}
\mathbf{r}(\hat{\tau}) = & \alpha \sqrt{P_\mathrm{T}} \mathbf{a}_\mathrm{R}(\theta, \phi) \mathbf{a}_\mathrm{T}^\text{T}(\theta, \phi) \mathbf{g}\nonumber \\ 
&\times\int_0^T \psi(t - \tau) \psi_{\text{MF}}^\text{*}(t - \hat{\tau}) \,\mathrm{d}t + \mathbf{n},
\end{align}
where $\mathbf{n}$ denotes the filtered noise. Assuming perfect synchronization, i.e., $\hat{\tau} = \tau$, the integral simplifies to one. The final receveid signal is denoted as
\begin{equation}
\mathbf{r}  = \alpha \mathbf{h}(\theta, \phi) + \mathbf{n},
\end{equation}
where $\mathbf{h}(\theta, \phi) = \sqrt{P_\mathrm{T}} \mathbf{a}_\mathrm{R}(\theta, \phi) \mathbf{a}_\mathrm{T}^\text{T}(\theta, \phi) \mathbf{g}.$ Perfect sampling is assumed following matched filtering to ensure aliasing-free signal representation, as justified by \cite{friedlander2012BeamMIMO}.

\section{Detection Problem Formulation}
The radar field of view is divided into $L_\mathrm{x} \times L_\mathrm{y}$ discrete angular bins, indexed by spatial frequencies $\nu_\mathrm{x}^{(i)}, i \in \{1, \dots, L_\mathrm{x}\}$, and $\nu_\mathrm{y}^{(j)}, j \in \{1, \dots, L_\mathrm{y}\}$.
 The radar transmits $K$ pulses, indexed by $k \in \{1, \dots, K\}$. For each spatial bin $m\in\mathcal{M}$ with $|\mathcal{M}| = L_\mathrm{x} L_\mathrm{y} $ (representing spatial location $(i,j)$) at time $k$, the received signal is modeled as
\begin{equation}
\mathbf{r}_{k,m} = \alpha_{k,m} \mathbf{h}_{k,m} + \mathbf{n}_{k,m},
\end{equation}
where \( \mathbf{n}_{k,m} \sim \mathcal{CN}(\mathbf{0}, \mathbf{\Gamma}) \) is a zero-mean complex Gaussian noise with covariance matrix \( \mathbf{\Gamma} \).
Target detection is framed as a binary hypothesis test
\begin{equation}
\begin{aligned}
\mathcal{H}_0 &: \quad \mathbf{r}_{k,m} = \mathbf{n}_{k,m}, \\
\mathcal{H}_1 &: \quad \mathbf{r}_{k,m} = \alpha_{k,m} \mathbf{h}_{k,m} + \mathbf{n}_{k,m},
\end{aligned}
\end{equation}
for all $k = 1, \dots, K$. Under $\mathcal{H}_0$, the observation contains only disturbance; under $\mathcal{H}_1$, a target signal is present in addition to the disturbance. To distinguish between $\mathcal{H}_0$ and $\mathcal{H}_1$, we use a test statistic, where we employ the CFAR adaptive matched filter detector. This detector is derived in a manner similar to the generalized likelihood ratio test (GLRT), but it uses a simplified test statistic that can be viewed as a limiting case of the GLRT detector~\cite{robey1992detector}. The test statistic is given by
\begin{equation}\label{eq:detector}
\Lambda_{k,m} =  \frac{\left| \mathbf{h}_{k,m}^\text{H} \mathbf{\Gamma}^{-1}\mathbf{r}_{k,m} \right|^2}
{\mathbf{h}_{k,m}^\text{H} \mathbf{\Gamma}^{-1}\mathbf{h}_{k,m}}\underset{\mathcal{H}_0}{\overset{\mathcal{H}_1}{\gtrless}} \eta.
\end{equation}
 In this work, we assume that the covariance matrix $\mathbf{\Gamma}$ is perfectly known. In practice, it is typically estimated from secondary data that is assumed to be free of signal components, i.e., collected under the null hypothesis $\mathcal{H}_0$. The threshold $\eta$ is chosen to guarantee a CFAR, and is given by $\eta = -\ln(P_\mathrm{FA}),$
where $P_\mathrm{FA}$ is the target false alarm probability~\cite{robey1992detector}.  
\section{RL-Based CR with TRIS Beamforming}
RL is a branch of machine learning where an agent learns by interacting with its environment~\cite{sutton2017RLIntro}. At each step $k$, it observes state $s_k$, takes action $a_k$, and receives reward $r_k$. Over time, it learns a policy $\pi$ mapping states to actions to maximize long-term rewards. Here, the radar is modeled as an RL agent that detects targets, adapts its strategy from states and rewards, and adjusts TRIS phase shifts based on learned target locations, as shown in Fig.~\ref{fig:system}.
\subsection{State Definition}
The state \(s_k\) is defined by the count of bins $m$ that exceed a threshold $\eta$, indicating target presence
\begin{equation} \label{eq:states}
    s_k = \sum_{m \in \mathcal{M}}\widehat{\Lambda}_{k,m},
\end{equation}
where $\widehat{\Lambda}_{k,m}$ is 1 if $\Lambda_{k,m} > \eta$, otherwise 0. The state space is defined by the maximum number of detectable targets $\bar{T}$, and is given by $\mathcal{S} = \{1, 2, \dots, \bar{T}\}$.
\subsection{Action Selection}
In the proposed system, the radar agent selects an action $a_k$ at each time step $k$ based on the current state $s_k$. Each action comprises two key components~\cite{ahmed2021RLradar}:
\begin{enumerate}
\item Identification of the angle bins most likely to contain targets based on the observed environmental state.  
\item Optimization of the phase shift matrix $\mathbf{\Phi}$ of the TRIS to steer the beampattern toward the selected bins, thereby maximizing the likelihood of target detection.
\end{enumerate}
In the first step the action $a_k$ is selected from the set of available actions $\mathcal{A} = \{\Theta_{\bar{t}} \mid \bar{t} \in \{0, 1, \dots, \bar{T}\}\}$, where $\Theta_{\bar{t}} \in \{1, \dots, \bar{t} \,\}$ is the set of $\bar{t}$ candidate angle bins. Each of these bins represents a potential target. The set $\Theta_{\bar{t}}$ is determined by selecting the $\bar{t}$ highest values of the detection metric $\Lambda_{k,m}$ in (\ref{eq:detector}). After identifying the set of bins $\Theta_k$ at time step $k$, the system then optimizes the phase shift matrix \(\mathbf{\Phi}\) to maximize the minimum beampattern gain across the selected bins. The problem is defined as
\begin{subequations}\label{eq:opt}
\begin{align}
        \max_{\mathbf{\Phi}, \, \gamma} \quad & \gamma \tag{\ref{eq:opt}} \\
        \text{s.t.} \quad 
        & B\!\left(\hat{\nu}_{x}^{(j)}, \hat{\nu}_{y}^{(j)}\right) \geq \gamma, &\forall j \in \Theta_p, \\
         &\varphi_n \in [0,2\pi),& n=1,\dots,N,
\end{align}
\end{subequations}
where $B(\hat{\nu}_{x}^{(j)}, \hat{\nu}_{y}^{(j)})$ is the beampattern of the chosen positions. Problem~\eqref{eq:opt} is a non-convex optimization due to the unit-modulus constraints on the phase shifts. To address this, we reformulate the problem via a Lagrangian relaxation and apply a successive optimization strategy inspired by majorization–minimization techniques~\cite{sun2017mm}.

\subsection{Reward Calculation}
The reward \(r_k\) reflects the detection performance and is calculated as
\begin{equation}\label{eq:reward}
    r_{k+1} = \sum_{m \in \mathcal{B}_\text{target}} \hat{P}_{D,{(m,k)}} - \sum_{m \in \mathcal{B}_\text{non-target}} \hat{P}_{D,{(m,k)}},
\end{equation}
where \(\mathcal{B}_\text{target}\) and \(\mathcal{B}_\text{non-target}\) represent bins with and without targets, respectively. $\hat{P}_{D,{(m,k)}}$ denotes the theoretical probability of detection~\cite{ahmed2021RLradar}.
\subsection{SARSA Algorithm for Detection}
In this work we use the SARSA algorithm to update the Q-function based on the sequence \(s_k, a_k, r_{k+1}, s_{k+1}, a_{k+1}\)~\cite{sutton2017RLIntro}. The SARSA process is presented in Algorithm~\ref{alg:alg1}, where the Q-function matrix \(\mathbf{Q}_{\bar{T}\times\bar{T}}\) approximates the cumulative reward from \(s_k\) and \(a_k\) under policy \(\pi\) using
\begin{align}\label{eq:qfunc}
    Q(s_k, a_k) \leftarrow & Q(s_k, a_k) +     \\
    &\alpha \left[ r_{k+1} + \gamma Q(s_{k+1}, a_{k+1}) - Q(s_k, a_k) \right], \nonumber
\end{align}
with learning rate \(\alpha\) and discount factor \(\gamma\). An \(\epsilon\)-greedy policy helps the radar agent choose actions to balance exploration and exploitation~\cite{sutton2017RLIntro}.
\begin{algorithm}[htb]
\caption{SARSA for Multi-Target Detection}
\label{alg:alg1}
\begin{algorithmic}[1]
    \State Initialize $\mathbf{Q} \gets 0_M$, $s_0 \gets 1$, $a_0 \gets 1$, and randomly initialize $\mathbf{\Phi}_k$
    \For{each time step $k$}
        \State Take action $a_k$ by transmitting $\mathbf{g}_k s(t)$ using $\mathbf{\Phi}_k$ 
        \State $\mathbf{r}_{k,m}\gets$ acquired received signal $\forall m$
        \State $s_{k+1}\gets$ result from (\ref{eq:states}), $r_{k+1}\gets$ result in (\ref{eq:reward})
        \State Choose $a_{k+1}$ using $\epsilon$-greedy, identify $\Theta_p$
        \State Update $Q(s_k, a_k)$ using (\ref{eq:qfunc})
        \State $s_k \gets s_{k+1}$; \quad $a_k \gets a_{k+1}$
        \If{$s_{k+1} \neq 0$}
            \State $\mathbf{\Phi}_{k+1}\gets$ solution of (\ref{eq:opt})
        \Else
            \State $\mathbf{\Phi}_{k+1}\gets$ random value
        \EndIf
    \EndFor
\end{algorithmic}
\end{algorithm}

\section{Simulation Results}
\begin{table}[h]
\centering
\caption{Simulation Parameters\label{tab:par}}%
\begin{tabular}{@{}ll@{}}
\toprule
Parameter  & Value \\
\midrule
Probability of false alarm $P_\mathrm{FA}$ & $10^{-4}$ \\ 
Power $P_\mathrm{T}$ & $1$  \\ 
Discount factor $\gamma$ & 0.8 \\
Greedy factor $\epsilon$ & 0.5 \\
Maximum number of targets $\Bar{T}$ & 10 \\
Elements size $d_x\times d_y$ & $\frac{\lambda}{2}\times\frac{\lambda}{2}$\\
Frequency $f$ & $28\,\text{GHz}$\\
Distance Antenna-RIS $d_L$~\cite{an2024SIM} & $20\lambda$ \\
\bottomrule
\end{tabular}
\end{table}
The simulations are performed on a grid consisting of $L_\mathrm{x} = L_\mathrm{y} = 20$ bins, yielding a total of $400$ bins. The grid is defined in the spatial frequency domain, where $\mathbf{\nu}_{x}$ and $\mathbf{\nu}_{y}$ stem from $\{-0.5, -0.45,\dots, 0.45\}$. Additional key parameters are listed in Table \ref{tab:par}.
\begin{figure}[h]
  \centering
  \centerline{\includegraphics[width=0.5\linewidth]{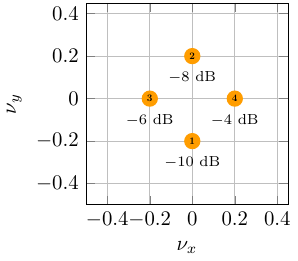}}
  \caption{Targets position and corresponding SNR}\label{fig:pos}
\end{figure}
We define a controlled scenario featuring four stationary targets positioned at fixed spatial coordinates. Each target maintains a constant SNR, creating a stable environment for evaluating baseline detection performance. The target locations and corresponding SNR's are shown in Figure~\ref{fig:pos}. The scenario is simulated over 500 Monte Carlo runs.
\begin{figure}[h]
  \centering
  \centerline{\includegraphics[width=1\linewidth]{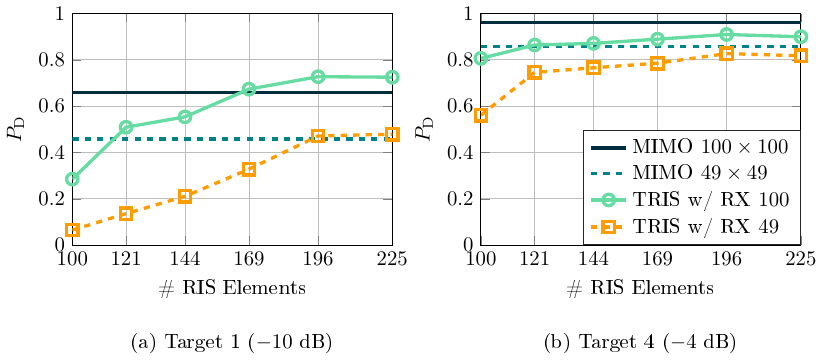}}
  \caption{Probability of detection for two TRIS configurations as a function of the number of elements, compared against two MIMO $\mathrm{TX} \times \mathrm{RX} $ configurations with $100\times100$ and $49\times49$.}\label{fig:tris}
\end{figure}
We compare the detection performance of a TRIS-aided CR with that of a conventional MIMO CR radar using RL~\cite{umra2025smartersensing2dclutter}. The MIMO radar employs a quadratic antenna array at both the transmitter and the receiver. We consider antenna array configurations of size $49 \times 49$ and $100 \times 100$, 
with an equal number of transmit ($\mathrm{TX}$) and receive ($\mathrm{RX}$) antennas. For the TRIS-aided radar, the receiver is equipped with $N_\mathrm{R} = 49$ or $N_\mathrm{R} = 100$ antennas, while the number of RIS elements is varied to assess its impact on detection performance. The MIMO CR employs full RF chains at each antenna, providing more degrees of freedom and full control of the transmit and receive apertures. In contrast, the TRIS-aided CR relies on passive RIS phase shifts and thus operates under more constrained conditions.

We now examine the simulation results in Figure~\ref{fig:tris}, which show the detection probabilities for target~1 and target~4, as target~1 corresponds to the lowest SNR and target~4 to the highest.
For target~1, characterized by a low SNR, the results in Figure~\ref{fig:tris}(a) indicate that the TRIS-aided radar attains a probability of detection  comparable to that of the MIMO system once the RIS is equipped with a sufficiently large number of reflecting elements. This can be attributed to the RIS effectively enhancing the received signal power through passive transmit beamforming, thereby compensating for the smaller number of physical antennas compared to the MIMO CR system. Moreover, when $N_\mathrm{R} = 100$, the TRIS configuration even surpasses the MIMO radar performance for more than $169$ RIS elements, demonstrating the scalability and effectiveness of the TRIS approach in low-SNR regimes. This improvement arises from the fact that, as the number of RIS elements increases, the generated beams become progressively narrower, thereby enhancing the focusing capability. Differently, for target~4, which operates at a relatively higher SNR as shown in Fig.~\ref{fig:tris}(b), the performance gap between TRIS-aided and MIMO radars narrows even faster with fewer elements compared to target~1, because high-SNR targets can still be reliably detected without requiring narrow beams. In both TRIS configurations, the $P_\mathrm{D}$ approaches that of the MIMO radar, and it is expected that with an even larger number of RIS elements, the TRIS radar would eventually achieve higher performance.

\section{Conclusion}
This paper presents a CR architecture that combines RL with TRIS for adaptive target detection. A SARSA agent optimizes TRIS phase shifts for beam steering and waveform shaping. Simulations show TRIS-aided radar, especially with larger arrays, outperforms MIMO RL in low-SNR regimes while reducing cost, energy, and size. Future work could investigate using TRIS at the receiver and evaluating performance in joint radar-communication scenarios, considering beamforming, interference management, and trade-offs between sensing and communication.
\clearpage
\bibliographystyle{IEEEbib}
\bibliography{bibliography} 
\end{document}